\documentstyle[aps]{revtex}

 \newcommand{\bq}{\begin{equation}}
 \newcommand{\eq}{\end{equation}}
 \newcommand{\bqn}{\begin{eqnarray}}
 \newcommand{\eqn}{\end{eqnarray}}
 \newcommand{\nb}{\nonumber}
 \newcommand{\lb}{\label}

\begin{document}

 \title{On Gauge Choice of  Spherically Symmetric  3-Branes}

\author{Anzhong Wang \thanks{E-mail: Anzhong$\_$Wang@baylor.edu} } 
\address{
Department of Physics, Zhejiang University of Technology, 
Hong Zhou, China\\  
and \\
CASPER, Department of Physics, Baylor University, Waco, TX76798}

\date{\today}

\maketitle

\begin{abstract}

Gauge choice for a spherically symmetric 3-brane embedded in a D-dimensional 
bulk with arbitrary matter fields on and off the brane is studied.  It is 
shown that Israel's junction conditions across the brane  restrict severely
the  dependence of the matter fields on the spacetime coordinates.  As examples, 
a scalar field or a Yang-Mills potential can be only either time-dependent  or 
radial-coordinate dependent for the chosen gauge, while for a perfect fluid 
it must be co-moving.

\end{abstract}

 \vspace{1.cm}


\maketitle

\vspace{1.cm}

\section{Introduction}

\renewcommand{\theequation}{1.\arabic{equation}}
\setcounter{equation}{0}

 A number of current unification theories such as string theory and M-theory
 suggest that we may live in a world that has more than three spatial
 dimensions.  Because only three of these are presently observable, one has to
 explain why the others are hidden from detection.  One such explanation
 is the so-called Kaluza-Klein (KK) compactification, according to which the
 size of the extra dimensions is very small (often taken to be on the order
 of the Planck length).  As a consequence, modes that have momentum in the
 directions of the extra dimensions are excited at currently inaccessible
 energies.

Recently, the braneworld scenario \cite{ADD,RS} has dramatically changed
 this point of view and, in the process, received a great deal of attention.
 At present, there are a number of proposed models  (See, for example,
 \cite{reviews} and references therein.).  In particular,
 Arkani-Hamed {\em et al} (ADD) \cite{ADD} pointed out that the extra
 dimensions need not necessarily be small and may even be on the scale of
 millimeters.  This model assumes that Standard Model fields are confined
 to a three (spatial) dimensional hypersurface (a 3-brane) living in a
 larger dimensional bulk space while the gravitational field propagates
 in the whole bulk.  Additional fields may live only on the brane or
 in the whole bulk, provided that their current undetectability is 
 consistent with experimental bounds.
 One of the most attractive features of this model is that it may
 potentially resolve the long standing hierarchy problem, namely the
 large difference in magnitudes
 between the Planck and electroweak scales.

In a different model, Randall and Sundrum (RS) \cite{RS}
 showed that if the self-gravity of the brane is included, gravitational
 effects can be localized near the brane at low energy and the 4D newtonian gravity
 will be reproduced on the brane even in the presence of infinitely large
 extra dimensions.  In this model, the $4D$ Planck scale, $M_{Pl}$, is
 determined by the curvature of the extra dimensions rather than by their size,
 as proposed in \cite{ADD}.

The RS model \footnote{In this Letter we are mainly concerned with the
so-called RS2 model, in which only one brane exists.} 
 was soon generalized to include arbitrary matter fields on the
 brane \cite{SMS00}.  In particular, Shiromizu, Maeda and Sasaki (SMS)
 considered the embedding of a 3-brane ($M, {}^{(4)}g_{AB}$) in a 5D bulk ($V,
 {}^{(5)}g_{AB}$), where the 3-brane metric is given by ${}^{(4)}g_{AB} =
 {}^{(5)}g_{AB} - n_{A}n_{B}$, and
 $n_{A}$ is the unit normal vector to the brane.  Note that
 we use Greek indices  to run from $0$ to $3$, uppercase Latin
 indices  to run from $0$ to $D-1$, and lowercase Latin indices to run
 from $0$ to $D-2$. Using the Gauss--Codacci
 relations, SMS wrote the 5D Einstein field equations ${}^{(5)}G_{AB} =
 \kappa^{2}_{5} \;{}^{(5)}T_{AB}$ in the form
 \bqn
 \lb{1.1a}
& &  {}^{(4)}R_{AB} - \frac{1}{2} \, {}^{(4)}g_{AB} \,
 {}^{(4)}R = {}^{(4)}{\cal{T}}_{AB},\\
 \lb{1.1b}
& &  D_{C}K^{C}_{A} - D_{A}K = \kappa^{2}_{5}\; {}^{(5)}T_{BC} \,
 {}^{(4)}g^{C}_{A} \, n^{B}, 
 \eqn
 where these equations are understood to apply in each of the two regions,
 $V^{+}(z \ge 0)$ and $V^{-}(z \le 0)$, and  $z$ denotes
the coordinate of the extra dimension and $z=0$ is the location of the brane. 
The quantity ${}^{(4)}{\cal{T}}_{AB}$ is given by
\bqn
\lb{1,1c}
{}^{(4)}{\cal{T}}_{AB} &\equiv& \frac{2{\kappa_{5}}^{2}}{3}
\left\{{}^{(5)}T_{CD}{}^{(4)}g^{C}_{A}\; {}^{(4)}g^{D}_{B} 
+ \left[{}^{(5)}T_{CD}n^{C}n^{D}
- \frac{1}{4}{}^{(5)}T^{C}_{C}\right]{}^{(4)}g_{AB}\right\}\nb\\
& & + K K_{AB} - K^{C}_{A}K_{BC} - \frac{1}{2}{}^{(4)}g_{AB}\left(K^{2}
- K^{CD}K_{CD}\right) - {\cal{E}}_{AB},\nb\\
{\cal{E}}_{AB} &\equiv& {}^{(5)}C^{E}_{FCD}n_{E} n^{C} \; {}^{(4)}g^{F}_{A} 
\; {}^{(4)}g^{D}_{B},
\eqn
and ${}^{(5)}C^{A}_{BCD}$ denotes the Weyl tensor of the bulk.

The boundary conditions at
 $z = 0$ are simply the Israel junction conditions \cite{Israel66},
 \bq
 \lb{1.4}
 \left[K_{AB}\right]^{-} = - \kappa^{2}_{5} \;
 \left(S_{AB} - \frac{1}{3} \, {}^{(4)} g_{AB} \, S\right),
 \eq
 where
 \bqn
\lb{1.2a}
\left[K_{AB}\right]^{-} &\equiv& \lim_{z \rightarrow 0^{+}}{K^{+}_{AB}}
                           - \lim_{z \rightarrow 0^{-}}{K^{-}_{AB}},\nb\\
\lb{1.2b}
S_{AB} &\equiv & {}^{(4)}T_{AB} - \lambda \, {}^{(4)}g_{AB},
\eqn
 with $\lambda$ and ${}^{(4)} T_{AB}$ being, respectively,  the cosmological constant
and the energy-momentum tensor on the 3-brane. 
 Combining the assumption of $Z_2$ symmetry with Eqs.~(\ref{1.4}) and
 (\ref{1.1a}) in the limit $z \rightarrow 0^{\pm}$,
 SMS obtained the effective 4D Einstein field equations on the 3-brane
\cite{SMS00}
 \bq
 \lb{1.3}
 {}^{(4)}G_{AB} = - \Lambda_{4} \, {}^{(4)}g_{AB}
                  + 8\pi G_{4}\; {}^{(4)}T_{AB}
                  + \kappa^{4}_{5} \, \pi_{AB}
                  - {\cal{E}}_{AB},
 \eq
where $G_{4} \equiv {\kappa_{5}}^{4}\lambda/(48\pi)$.
These equations include two correction terms to the original Einstein
 equations, namely
 ${\cal{E}}_{AB}$ and $\pi_{AB}$.
 In the weak field limit, the first term gives rise to the massive KK
 modes of the graviton \cite{RS}.
 The second term, $\pi_{AB}$,
 is negligible for weak fields.  However, in strong field
 situations, such as those at the
 threshold of black hole formation, it is expected to play a
 significant role.  Indeed, it would appear to be this term which is the origin
 of the result of \cite{Bru01} that, given the 4D projected Einstein equations
 (\ref{1.3}) and the Israel matching conditions (\ref{1.4}), {\em  the
 exterior of a collapsing homogeneous dust cloud cannot be
 static}.  This  result was further generalized to other cases \cite{GD02},
and represents a significant departure from the familiar
 result in Einstein's theory, in which the vacuum exterior must be the
 static Schwarschild solution.  A number of other recent results suggest
 additional phenomena different from the standard predictions of general
 relativity \cite{Na01}. In particular, it was argued that static
 braneworld black holes might not exist at all \cite{Ta02}.

Braneworld scenarios have further been promoted by the possibility that they
may provide the origin of dark energy \cite{BraneDE}, which is needed in order
to explain why our universe is currently accelerating \cite{DE}.

In this Letter, we report on some results  concerning the gauge choice 
for  matter fields confined on a  spherically symmetric 3-brane.  
We show that  the boundary conditions 
(\ref{1.4}) serve as very strong restrictions on the possible dependence 
of the matter fields on the spacetime coordinates.  In particular, 
for a particular choice of the gauge, a scalar field  or
a  Yang-Mills field can be only either time-dependent or radial-coordinate 
dependent, while for a perfect fluid its radial velocity must vanish.
Moreover, these conclusions would appear to be true
not only for the generalized RS models in a 5D bulk, but
also for 3-branes in higher dimensional spacetimes \cite{reviews}.
This is quite different from its four-dimensional counterpart.

Before showing these results, let us first give a brief review on the gauge 
choice of a 4D spacetime with spherical symmetry, for which the general 
metric can be cast, without loss of  generality, in the form,
\bq
\lb{1.5}
ds^{2}_{4} = g_{ab}\left(x^{c}\right)dx^{a} dx^{b}
+ s^{2}\left(x^{c}\right) d\Omega^{2},\; (a,b , c = 0, 1),
\eq
where $d\Omega^{2} \equiv d\theta^{2} + \sin^{2}\theta d\varphi^{2}$.
Clearly, the form of the metric is invariant under the coordinate 
transformations,
\bq
\lb{1.6}
x^{0} = x^{0}\left({x'}^{0}, {x'}^{1}\right),\;\;\;
x^{1} = x^{1}\left({x'}^{0}, {x'}^{1}\right).
\eq
Using these two degrees of the gauge freedom, one can choose different 
gauges for different
matter fields. Because of the complexity of the Einstein field equations,
such choices often are crucial in studying the problem. For example,
for a collapsing perfect fluid, one usually chooses the so-called comoving
gauge,
\bq
\lb{1.7}
g_{01}\left(x^{c}\right) = 0, \;\;\;
u_{A} = \left(-g_{00}\right)^{1/2} \delta^{0}_{A},
\eq
so that the metric takes the form,
\bq
\lb{metric1}
ds^{2}_{4} = g_{00}\left(x^{c}\right)\left(dx^{0}\right)^{2}
+  g_{11}\left(x^{c}\right)\left(dx^{1}\right)^{2}
+ s^{2}\left(x^{c}\right) d\Omega^{2},\; ( c = 0, 1),
\eq
where $u_{A}$ denotes the four-velocity of the fluid, and $x^{0}$ is the 
time-like coordinate. For a collapsing scalar field, on the othe hand,
 a possible choice of the gauge is
\bq
\lb{1.8}
g_{01}\left(x^{c}\right) = 0, \;\;\;
s\left(x^{c}\right) = x^{1} \equiv  r,
\eq
so that the metric takes the form,
\bq
\lb{metric2}
ds^{2}_{4} = g_{00}\left(t,r\right)dt^{2}
+  g_{11}\left(t,r\right)dr^{2}
+ r^{2} d\Omega^{2},
\eq
where $ t \equiv  x^{0}$ and $ \phi = \phi(t, r)$, with $\phi$ denoting
the scalar field. Certainly, depending on a specific problem to be considered, 
other  gauges can be chosen.

\section{Gauge Freedom and Gauge Choice For a Spherical 3-Brane}

\renewcommand{\theequation}{2.\arabic{equation}}
\setcounter{equation}{0}

To begin with, consider the general action
 describing a 3-brane embedded in a D--dimensional bulk \cite{Note},
 \bqn
 \lb{2.1}
 S &=&
 \frac{1}{2\kappa^{2}_{D}}\int_{M_{D}}{d^{D}x\sqrt{-{}^{(D)}g}\left({}^{(D)}R -
 2\Lambda_{D} + 2\kappa^{2}_{D}{\cal{L}}^{B}_{m}\right)}\nb\\
 & & + \frac{1}{2\kappa^{2}}\int_{M_{4}}{d^{4}x\sqrt{-g}
 \left(- 2\Lambda + 2\kappa^{2}{\cal{L}}_{m}\right)},
 \eqn
 where ${}^{(D)}R,\; \Lambda_{D}$ and ${\cal{L}}^{B}_{m}$ ($R,\; \Lambda,\;
 {\cal{L}}_{m}$) denote, respectively, the Ricci scalar, cosmological constant
 and matter content of the bulk (of the 3-brane). The constants $\kappa_{D}$
 and $\kappa$ are related to the Planck scales $M$ and $M_{Pl}$, respectively,
 by 
 \bqn
 \lb{2.1a}
 \kappa^{2}_{D} &=& 8\pi {}^{(D)}G = M^{2 - D},\nb\\
 \kappa^{2} &=& 8\pi  G = M^{-2}_{Pl}, 
 \eqn
 with $D \equiv  n + 4$.
 The bulk metric is ${}^{(D)}g_{MN}$ and $g_{\mu\nu}$ denotes
 the induced metric on the 3-brane, located on the surface  
 $\Phi(x^{A}) = 0$.  Taking a different approach from SMS, we vary
 Eq.~(\ref{2.1}) with respect to ${}^{(D)}g_{MN}$ and $g_{\mu\nu}$ to get
 the full D--dimensional Einstein field equations in the form
 \bqn
 \lb{2.3a}
 {}^{(D)}G_{MN}^{+} &=& \kappa_{D}^{2} T^{B +}_{MN}
  - \Lambda_{D} {}^{(D)}g_{MN}^{+},\; (\Phi \ge 0),\\
 \lb{2.3b}
 {}^{(D)}G_{MN}^{-} &=& \kappa_{D}^{2} T^{B -}_{MN}
  - \Lambda_{D} {}^{(D)}g_{MN}^{-},\; (\Phi \le 0),\\
 \lb{2.3c}
 {}^{(4)}G_{\mu\nu}^{Im} &=& \kappa_{D}^{2}\left(T_{\mu\nu} -
 \frac{\Lambda}{\kappa^{2}}g_{\mu\nu}\right),\; (\Phi = 0),
 \eqn
 where $T^{B}_{MN}$ and $T_{\mu\nu}$ denote the stress energy
 tensor of the bulk and of the
 3-brane, respectively. The quantities with superscript ``$+$" (``$-$") denote
 those calculated in the region $\Phi \ge 0$ ($\Phi \le 0$), and
 ${}^{(4)}G_{\mu\nu}^{Im}$ denotes the delta-function-like (impulsive) part of
 $G_{MN}$ with support on the 3-brane,
\bq
\lb{einstein}
 {}^{(D)}G_{MN} =  {}^{(D)}G_{MN}^{+}H\left(\Phi\right)
+ {}^{(D)}G_{MN}^{-}\left[1 - H\left(\Phi\right)\right]
+  {}^{(4)}G_{\mu\nu}^{Im} \delta^{\mu}_{M} \delta^{\nu}_{N}
\delta\left(\Phi\right),
\eq
where $ \delta\left(\Phi\right)$ denotes the Dirac delta function,
and  $H(\Phi)$  the Heavside function, defined as
\bq
\lb{heavside}
 H\left(\Phi\right) = \cases{1, & $ \Phi \ge  0$,\cr
0, & $ \Phi <  0$.\cr}
\eq
Eq.(\ref{einstein}) can easily be obtained by the following considerations.
Let us first denote the region with $\Phi \ge 0$ as   $V^{+}$, 
the region with $\Phi \le 0$ as   $V^{-}$, and the hypersurface $\Phi = 0$ 
as  $\Sigma$. Then, since the Einstein field equations involve the second-order 
derivatives of the metric coefficients, one can see that the metric must be at 
least $C^{2}$ in regions $V^{\pm}$ and $C^{0}$ across the hypersurface $\Phi = 0$, 
so that the Einstein field equations (or any second-order differential equations
involved) 
hold in the sense of distributions  \cite{Taub80}. Consequently, the metric 
$g_{AB}$ in the whole spacetime can be written as
\bq
\lb{2.3bb}
g_{AB} = g^{+}_{AB}H\left(\Phi\right) 
     + g^{-}_{AB}\left[1 - H\left(\Phi\right)\right],
\eq
where  quantities with superscripts $``+"$ ($``-"$) denote the ones 
defined in $V^{+}$ ($V^{-}$). Hence, we find that
\bqn
\lb{2.3cc}
g_{AB,C} &=& g^{+}_{AB,C}H\left(\Phi\right) 
           + g^{-}_{AB,C}\left[1 - H\left(\Phi\right)\right],\nb\\
g_{AB,CD} &=& g^{+}_{AB,CD}H\left(\Phi\right) 
           + g^{-}_{AB,CD}\left[1 - H\left(\Phi\right)\right]
	   + \left[g_{AB,C}\right]^{-} \Phi_{,D}\delta\left(\Phi\right),\nb\\
\eqn
where $(\;)_{,C} \equiv \partial(\;)/\partial x^{C}$ and
\bq
\lb{2.3d}
\left[g_{AB,C}\right]^{-} \equiv \lim_{\Phi \rightarrow 0^{+}} 
{\frac{\partial{g_{AB}^{+}}} {\partial x^{C}}} - 
\lim_{\Phi \rightarrow 0^{-}} 
{\frac{\partial{g_{AB}^{-}}} {\partial x^{C}}}.
\eq
From Eq.(\ref{2.3cc}) and the following,  
\bqn 
\lb{2.3e}
& & H^{m}\left(\Phi\right) = H\left(\Phi\right),\;\;\;
\left[1 - H\left(\Phi\right)\right]^{m} = 1 - H\left(\Phi\right),\nb\\
& & H\left(\Phi\right)\left[1 - H\left(\Phi\right)\right] = 0, \;\;\;
\left[1 - H\left(\Phi\right)\right] \delta(\Phi) 
= \frac{1}{2} \delta(\Phi) =  H\left(\Phi\right)\delta(\Phi),
\eqn
where $m$ is an integer,   we can easily see that the Einstein tensor 
$G_{AB}$ can be  written, in general, in the form of Eq.(\ref{einstein}).

In the 5D case a 3-brane is a
 hypersurface of a 5D bulk, and one can show that  Eq.~(\ref{2.3c}) is
 identical to Eq.~(\ref{1.4}) when written out in terms of the extrinsic
 curvature of the 3-brane \cite{Taub80}. Similarly, using the
 Gauss--Codacci relations one can show that Eqs.~(\ref{1.1a}) and
 (\ref{1.1b}) follow from Eqs.~(\ref{2.3a}) and (\ref{2.3b}).  It is, at the
 same time, worth emphasizing that Eqs.~(\ref{2.3a}--\ref{2.3b})
 contain additional information not present in Eqs.~(\ref{1.1a}--\ref{1.1b}).
 This includes, for instance, information about the evolution of
 $\cal{E}_{AB}$.
 For the case that $D \ge 6$, a 3-brane is a surface of co--dimension
 $(D-4)$ with respect to the bulk, and the problem becomes more subtle.  In
 particular, the generalization of the Gauss--Codacci
 relations and the Israel junction conditions to these cases
 has not, to our knowledge, been worked out.  For this reason, in the
 $D \ge 6$ cases we will only consider models with additional symmetries.  In
 this way, a 3-brane can be considered as a degenerate hypersurface.
 Indeed, this assumption turns out to include most of the braneworld models with
 higher dimensional bulks which have been studied so far \cite{reviews}.

\subsection{$ D = 5$}

Considering first the case $D = 5$, the most general bulk metric
 with a $S^{2}$ symmetry takes the form,
 \bq
 \lb{2.4}
 ds^{2} = g_{ij}\left(x^{k}\right)dx^{i}dx^{j}
         + s^{2}\left(x^{k}\right)d\Omega^{2},
 \eq
 where  $i$,  $\,j$ and $k$ are taken here to range over $0$, 
 $\, 1$ and $2$ with $z \equiv x^{2}$.  The location of a 3-brane 
 with spherical symmetry in general can be written as 
\bq
\lb{2.4a}
\Phi\left(x^0, x^1, z\right) = 0.
\eq

Note that the form of the metric (\ref{2.4}) is invariant under the
 coordinate transformations, 
\bq
\lb{2.4b}
x^{i} =  f^{i}\left(\bar{x}^{j}\right),\; (i, j = 0, 1, 2).
\eq
 As a result, using these three degrees of   freedom,
we can choose coordinates
 such that the brane is always located on the hypersurface $z = 0$ and
 \bq
 \lb{2.4c}
 \Phi\left(x^{0}, x^{1}, z\right) = z, 
 \;\;\; g_{0 z}(x^0, x^1, z) = g_{1 z}(x^0, x^1, z) = 0.
 \eq
 This choice of coordinates will be referred to as the {\em canonical
 gauge}.

Using this form for the metric together with the definition of the
 extrinsic curvature, $K_{AB} \equiv
 h_{A}{}^{C} h_{B}{}^{D} \nabla_{C} n_{D}$, we find that
 \bq
\lb{2.4d}
K_{\mu\nu} = (2N)^{-1}\frac{\partial g_{\mu\nu}(x^{\alpha},z)}
{\partial z},
\eq
 where  $n_{A} = N \delta^{z}_{A},\; N \equiv \sqrt{g_{zz}}$, and
 $\nabla_{C}$ denotes the covariant derivative with respect to ${}^{(D)}g_{AB}$.
 For the metric (\ref{2.4}) in the canonical gauge  we find that
 the Israel junction conditions, (\ref{1.4}), yield
 \bq
 \lb{2.6}
 T_{\mu\nu} =   \lambda g_{\mu\nu}
              - \frac{1}{2\kappa^{2}_{5}N}
                   \left(   \left[g_{\mu\nu,z}\right]^{-}
                          - g_{\mu\nu} g^{\alpha\beta}
                                  \left[g_{\alpha\beta,z}\right]^{-}
                  \right).
 \eq

It should be noted that even in the canonical gauge, there is residual 
coordinate freedom { on} the 3-brane:  
\bq
\lb{2.6a}
x^0 = F^{0}(\bar{x}^0, \bar{x}^1),\;\;\;
 x^1 = F^{1}(\bar{x}^0, \bar{x}^1).
\eq
We can exploit this remaining freedom and set
 \bq
 \lb{2.4e}
 g_{01}(x^0, x^1, 0) = g_{01,z}(x^0, x^1, 0) = 0,
 \eq
 so that the reduced metric  on the 3-brane takes the form,
\bq
\lb{reducedmetric}
\left. ds^{2}\right|_{z = 0} = 
\gamma_{00}(x^0, x^1)\left(dx^{0}\right)^{2}
+ \gamma_{11}(x^0, x^1)\left(dx^{1}\right)^{2} 
+ \gamma_{22}(x^0, x^1)d\Omega^{2},
\eq
where $\gamma_{ab}(x^0, x^1)  \equiv  g_{ab}(x^0, x^1, 0)$. It is remarkable
to note that {\em Eq.(\ref{2.4e}) is possible only in the cases where one of
the three energy conditions, weak, strong and dominant, holds} \cite{HE73}. 
To show this, 
following Chandrasekhar  \cite{Chandra83},  we first
transform to the coordinates $\bar{t}$ and $\bar{r}$  using the transformations
$x^0 = \phi(\bar{t},\bar{r})$ and $x^1 = \psi(\bar{t},\bar{r})$, in which the 
metric takes the form
\bq
\lb{2.6bb}
 ds^2 = -b \, d\bar{t}^2 + 2 c \, d\bar{t} d\bar{r} + d \, d\bar{r}^2
         + s^2  d\Omega^2  + N^2 dz^2,
\eq
 where  $b$, $\, c$, $d$, and $s$  are functions of
 $\bar{t},\; \bar{r}$ and $z$, and  have
 the properties   \cite{Chandra83},
\bq
\lb{gauge1}
 b(\bar{t},\bar{r},0) = d(\bar{t},\bar{r},0),\;\;\;
 c(\bar{t},\bar{r},0) = 0,
\eq
 at $z=0$. If we now make another coordinate transformation
 $\bar{t} = \Phi(t,r),\; \bar{r} = \Psi(t,r)$, it is straightforward to
 show that the conditions $g_{tr}(t, r, 0) = 0$ and $g_{tr,z}(t, r, 0) = 0$
 reduce to
\bqn
\lb{eq1}
 & & \Phi_{,t}  =  A \, \Psi_{,t} , \\
\lb{eq2}
& &  \Phi_{,r}   =   A^{-1} \, \Psi_{,r} ,\\
\lb{eq3}
 & & (A \Psi_{,t})_{,r}  =  (A^{-1} \, \Psi_{,r})_{,t},
 \eqn
 where
\bqn
\lb{ss}
A(t,r) &\equiv& \frac{1}{2 c_{,z}} \left[(b_{,z} - d_{,z}) 
\pm \Delta^{1/2}\right]_{z=0},\nb\\
\Delta(t,r) & \equiv & (b_{,z} - d_{,z})^2 - 4 c_{,z}{}^2 \, \vert_{z=0}.
\eqn
Eq.(\ref{eq3}) represents  the integrability condition of  Eqs.(\ref{eq1})
and (\ref{eq2}). 
Since  the metric coefficients is  at least $C^{2}$
in regions $V^{\pm}$ and $C^{0}$ across the hypersurface $z = 0$,  we can see
that, with respect to $t$ and $r$, the metric is  also at least 
$C^{2}$ even across  the hypersurface  $z = 0$. For such a  $C^{2}$ 
 metric, the theorems given 
in  \cite{Bernstein} show that  there will always exist a region of the 
$(t,r)$-plane for which Eqs.(\ref{eq1})-(\ref{eq3})  have solutions.  However,
 such solutions will be real only if 
 \bq
 \lb{sd}
 \Delta \ge 0.
 \eq
 This condition is ensured by assuming any of the standard energy conditions
 \cite{HE73}.  
 Indeed, using the Israel junction conditions (\ref{1.4}), we find
 \bq
 \Delta = (\kappa_5 b N)^2 \left[ (\rho + p_{\bar{r}})^2 - 4 q^2 \right],
 \eq
 where 
\bq
\lb{ss1}
\rho \equiv \frac{ T_{\bar{t}\,\bar{t}}}{b},\;\;\;
 p_{\bar{r}} \equiv \frac{T_{\bar{r}\,\bar{r}}}{d},\;\;\;
q \equiv \frac{T_{\bar{t}\,\bar{r}}}{(bd)^{1/2}}. 
\eq
 As shown in \cite{KST88},
{\em  a necessary condition for any of the weak, dominant, and strong
 energy conditions  to hold is  $\Delta \ge 0$}.
 We thus conclude that {\em a coordinate transformation exists such that 
(\ref{2.4e}) holds, provided that one of the three energy conditions holds}.

As a consequence of our coordinates and Eq.~(\ref{2.6}),
 $T_{\mu\nu}$ must  be diagonal. In particular, we have 
\bq
\lb{diagonal}
T_{tr}(t, r) = 0.
\eq
This  represents a very strong restriction on the  dependence of matter fields
confined to the 3-brane on the spacetime coordinates.   
To see this clearly, let us first consider   a scalar field
 $\phi$, for which the stress tensor is given by
\bq
\lb{aa}
T^{\phi}_{\mu\nu} =  \nabla_{\mu}\phi\nabla_{\nu}\phi 
- \frac{1}{2}g_{\mu\nu}[(\nabla\phi)^{2} + 2V(\phi)].
\eq
 In this case, we have 
\bq
\lb{bb}
\Delta = b^{-2} ( \phi_{t}{}^2 - \phi_{r}{}^2 )^2,\;\;\;
T^{\phi}_{tr}(t, r) = \phi_{,t}(t, r)\phi_{,r}(t, r).
\eq
Then, Eq.(\ref{diagonal})  implies   
\bq
\lb{cc}
(i) \;\; \phi(t, r) = \phi(t),\;\;\; {\mbox{or}} \;\;\;
(ii) \;\; \phi(t, r) = \phi(r).
\eq
 One can show that this is also true for a spherically
 symmetric $SU(2)$ Yang-Mills field.  In that case  the relevant term is
\bq
\lb{dd}
T^{\rm YM}_{tr} \propto w_{,t}(t, r)w_{,r}(t, r),
\eq
 where $w$ is the Yang-Mills potential \cite{Zhou92}.

Similarly, if one considers   a perfect fluid with stress tensor
\bq
\lb{ee}
T^{\rm fl}_{\mu\nu} = (\rho + p) u_\mu u_\nu + p \, g_{\mu\nu},
\eq 
for which 
\bq
\lb{ff}
\Delta = (\rho + p)^2,\;\;\; T^{\rm fl}_{tr} = (\rho + p) u_{t} u_{r},
\eq
it turns out that the fluid cannot have a radial velocity, that is, for the
present choice of the gauge, we must have
\bq
\lb{gg}
u_{r} = 0.
\eq 
It is interesting to note that the condition (\ref{sd}) is satisfied for the 
cosmological constant, for which the corresponding energy-momentum tensor is
given by Eq.(\ref{ee}) with $p = -\rho = \lambda$. 

The above results hold not only for $D = 5$ but also for $D \ge 6$. To see this,
in the following let us first consider the case where the 3-brane is 
a surface of co--dimension two, that is, $D = 6$.  Then, we shall further
generalize our results to the case $D > 6$.

\subsection{$D =  6$}

 Because we now have two extra spatial
 directions, the generalization of the RS model to include matter fields
 on the 3-brane becomes non-trivial.  Following Israel \cite{Israel77},
 we will assume a cylindrical
 symmetry in these extra dimensions and that the 3-brane is located on the
 4-dimensional surface $\rho = 0$ where $\rho$ and $\psi$ are chosen as
 polar-like coordinates for the two extra dimensions, and $\rho = 0$ is
 the symmetry axis for the cylindrical symmetry.
 This includes most of the braneworld models in six dimensional spacetimes
 \cite{reviews}.

Under these assumptions, it can be shown that the general bulk metric with a
 spherical 3-brane can be cast in the form
 \bqn
 \lb{2.8}
 ds^{2} &=&  - \alpha^{2}dt^{2} + a^{2} (dr + \beta dt)^{2} 
+ s^{2}d\Omega^{2}\nb\\
& &  + N^{2}d\rho^{2} + f^{2}\rho^{2}\left(d\psi + \omega dt\right)^{2} +
 g^{2} \rho^{2} d\psi^{2},
 \eqn
 where $\omega$ represents the rotation of the 3-brane, and all the metric
 coefficients are functions of $t,\; r$ and $\rho$, subject to the gauge
 (\ref{2.4e}).   Because the symmetry axis now
 represents a  3-brane, certain conditions must be imposed there \cite{PW00}.
 For the present purpose, it is sufficient to assume that:  (a) the symmetry
 axis must exist, that is,
 \bq
 \lb{cd1}
 X \equiv \left|\left|\partial_{\psi}\right|\right| \rightarrow O(\rho^{2}),
 \eq
as $\rho \rightarrow 0$, where $\partial_{\psi}$ is
 the cylindrical Killing vector with closed orbits, and that (b) the
 spacetime is free of curvature singularities on the axis, which can be
 assured by assuming the local flatness condition,
 \bq
 \lb{cd2}
\lim_{\rho \rightarrow 0^{+}}
         \frac{X_{,A}X_{,B} \, {}^{(D)}g^{AB}}{4X}  = 1.
\eq

To generalize Israel's method \cite{Israel66} to this case, we first
 calculate the extrinsic curvature $K_{ab}$ of the hypersurface $\rho =
 \epsilon$ and then take the limit $\epsilon \rightarrow 0$.  Introducing the
 quantities ${\cal{K}}_{ab}$ by 
\bq
\lb{curvature}
{\cal{K}}_{ab} \equiv \lim_{\rho \rightarrow 0^{+}}\left(\sqrt{-
 {}^{(5)}g} \, K_{ab}\right),
\eq
 the surface stress energy tensor can be defined as \cite{Israel77}
 \bq
 \lb{2.10}
 T^{a}_{b} = - \left({\cal{K}}^{a}_{b}
 - \delta^{a}_{b}{\cal{K}}^{c}_{c}\right),
 \eq
 provided ${\cal{K}}^{a}_{[c}{\cal{K}}^{b}_{d]} \not= 0$ and
 where we have let $a, \; b = 0, \cdots, 4$ and $x^{5} = \rho$.
 In this case, it can be shown that the extrinsic curvature of the
 hypersurface $\rho = \epsilon$ is given by
 \bq
 \lb{2.5a}
 K_{ab} = \frac{1}{2N}
 \frac{\partial g_{ab}(x^{c},\epsilon)}{\partial \rho}.
 \eq

For the case ${\cal{K}}^{a}_{[c}{\cal{K}}^{b}_{d]} = 0$, the surface
 stress energy tensor is instead defined as \cite{Israel77},
 \bq
 \lb{2.11}
 T^{a}_{b} = \frac{2\pi}{N^{2}\sqrt{f^{2} + g^{2}}}
 \left(\sqrt{f^{2} + g^{2}} - N\right) \delta^{a}_{\mu}\delta^{\nu}_{b},
 \eq
 where $\mu, \nu = 0, \cdots 3$.
 In passing, we note this case also corresponds to a cosmic string in
 a 6D bulk \cite{Gr00}.

However, in each of the above two cases it can be seen that the condition
Eq.(\ref{diagonal})   holds. Therefore, the junction conditions across
the 3-brane in a 6D bulk yield the same restrictions on the dependence
of the spacetime coordinates of the matter fields confined on the 3-brane
as those in the 5D case.

\subsection{$D >  6$}

When $D >  6$, we consider only the case where the extra $n$--dimensional
 space has an $SO(n-1)$ symmetry so that the bulk metric can be written in
 the form,
 \bqn
 \lb{2.13}
 ds^{2} & = &  - \alpha^{2}dt^{2} + a^{2} ( dr + \beta dt )^{2} 
+ s^{2}d\Omega^{2}\nb\\
& & +  N^{2}\left(d\rho^{2} + \rho^{2}d\Omega_{n-1}^{2}\right),
 \eqn
 where $d\Omega_{n-1}^{2}$ denotes the metric of a unit $(n-1)$--dimensional
 sphere, and all the metric coefficients are functions of $t, r$ and $\rho$.
 Using, as before, the coordinate freedom $t = F_{1}(t', r') $ and 
$r = F_{2}(t', r')$
 we can always assume that Eq.~(\ref{2.4e}) holds.
 From Eq.~(\ref{2.13}) we note that $\rho = 0$ represents
 a four-dimensional spacetime with spherical symmetry.  This we shall take as
 our 3-brane. Certainly, this is acceptable only after some (physical and or
 geometrical)
 conditions are satisfied at $\rho = 0$.  These will include, as before,
 that the
 spacetime is free of curvature singularities there.  In order to generalize
 Israel's
 method to this case, we also require that the limit
\bq
\lb{sssa}
{\cal{K}}_{ab} = \lim_{\rho \rightarrow 0^{+}}\left(\sqrt{-
 {}^{(D-1)}g} \, K_{ab}\right),
\eq
 exists, where $K_{ab}$ denotes the extrinsic
 curvature of the hypersurface $\rho = \epsilon$, but now with $a, b = 0, 1,
 \cdots, D-2$, and $x^{D-1} = \rho$.  With these conditions, we can
 define the surface stress energy tensor as that given by Eqs.~(\ref{2.10})
 and (\ref{2.11}).
 For the metric (\ref{2.13}), it can be shown that the extrinsic curvature
 $K_{ab}$ of the hypersurface $\rho = \epsilon$ is also given by
 Eq.~(\ref{2.5a}). Substituting it into Eq.~(\ref{2.10}), we find again that the
 component $T_{tr}(t, r)$ vanishes identically for both of the cases described
 by Eqs.~(\ref{2.10}) and (\ref{2.11}). Thus, the same restrictions
 on the dependence of the spacetime coordinates of the matter fields
confined to the 3-brane that  occur in the 5D case continue to hold  for a
 D--dimensional bulk given by metric (\ref{2.13}).

\section{Conclusions}

\renewcommand{\theequation}{3.\arabic{equation}}
\setcounter{equation}{0}

In summary, we have studied the embedding of a spherically
symmetric 3-brane into a D--dimensional bulk with arbitrary matter fields both
on the brane and in the bulk in the context of the braneworld scenario. We
have found that, for a particular choice of gauge,
the boundary (Israel's junction) conditions across the brane
together with imposition of the weak energy condition
provide very strong restrictions on the dependence of matter fields 
confined to the 3-brane 
on the spacetime coordinates.  As examples, a scalar field or a Yang-Mills
field can be only either time-dependent or radial-coordinate dependent,
while for a perfect fluid its radial velocity must vanish.

In this paper, we have studied only the function dependence of the metric 
coefficients for the canonical gauge. It would be very interesting to study the
effects of the matter fields in the bulk on the effective 4-dimensional
energy-momentum tensor ${}^{(4)}T_{AB}$, a subject that is under our current
investigation. Another interesting problem is the applications of the results 
obtained in this paper to cosmology \cite{Zhuk}.

\section*{\bf Acknowledgments}

The author would like to thank Rong-Gen Cai, Andrew Chamblin, Michael 
Christensen, Roy Maartens,  and Zhong-Chao Wu for useful conversations 
and comments.  His special thanks goes to E.W. Hirschmann for his valuable  
collaboration in the early stage of this work.  Part of the work was 
done when  the author was visiting Physics Department of Brigham Young 
University (BYU), and the Astrophysics Center, Zhejiang University of 
Technology (ZUT). He  would like to express his gratitude to them 
for their hospitality. The financial  assistance from Baylor University 
for the 2005 summer sabbatical leave, BYU, and ZUT  is gratefully 
acknowledged.


\begin{thebibliography}{99}


 \bibitem{ADD}  N. Arkani-Hamed, S. Dimopoulos and G. Dvali, Phys. Lett. {\bf
 B429}, 263 (1998); Phys. Rev. {\bf D59}, 086004 (1999); I. Antoniadis, N.
 Arkani-Hamed, S. Dimopoulos and G. Dvali, Phys. Lett., {\bf B436},  257 (1998).

\bibitem{RS} L. Randall and  R. Sundrum, Phys. Rev. Lett. {\bf 83},
3370  (1999), {\em ibid.},  {\bf 83}, 4690  (1999).

\bibitem{reviews} V.A. Rubakov,  Phys. Usp. {\bf 44}, 871 (2001); S. F\"orste,
  Fortsch. Phys. {\bf 50},  221 (2002); C.P. Burgess, {\em et al}, JHEP, {\bf
 0201}, 014 (2002); E. Papantonopoulos,  Lect. Notes Phys.  {\bf 592}, 458 (2002);
R. Maartens,  Living Reviews of Relativity {\bf 7} (2004); U. G\"unther and A. Zhuk,
``{\em Phenomenology of brane-World Cosmological Models,}" arXiv:gr-qc/0410130
(2004).

\bibitem{SMS00} T. Shiromizu, K.I. Maeda, and M. Sasaki, Phys. Rev. {\bf D62},
 024012 (2000).

\bibitem{Israel66} W. Israel, Nuovo Cim. {\bf 44B}, 1 (1966); {\em ibid.},
 {\bf 48B}, 463 (1967).


 \bibitem{Bru01} M. Bruni, C. Germani, and R. Maartens, Phys. Rev. Lett. {\bf
 87}, 231302-1 (2001).

 \bibitem{GD02} M. Govender and N. Dadhich, Phys. Lett. {\bf B538}, 233 (2002);
G. Kofinas and E. Papantonopoulos,  JCAP, {\bf 0412},  011 (2004).

 

\bibitem{Na01} K.I. Nakao, K. Nakamura, and T. Mishima,Phys. Lett. {\bf B564}, 
143 (2003); D. Ida  and K.-I. Nakao, Phys. Rev. {\bf D66}, 064026 (2002);
 Y. Shtanov and V. Sahni, Class. Quantum Grav. {\bf 19}, L101 (2002).
 (2002).

\bibitem{Ta02} T. Tanaka,  Prog. Theor. Phys. Suppl. {\bf 148}, 307 (2003).

\bibitem{BraneDE}  P.J. Peebles and A. Vilenkin, Phys. Rev.  {\bf D59}, 063505 (1999); 
%
R. Maartens, D. Wands, B.A. Bassett and I.P.C. heard, {\em ibid.}, {\bf 62}, 041301(R) (2000); 
%
G.~Dvali, G.~Gabadadze, and M.~Porrati, Phys. Lett. {\bf B485}, 208 (2000); 
%
E.J. Copeland, A. R. Liddle and J. E. Lidsey, Phys. Rev.  {\bf D64}, 023509 (2001); 
%
A. S. Majumdar, {\em ibid.}, {\bf 64}, 083503 (2001);
%
G. Huey and Lidsey, Phys. Lett.  {\bf B514}, 217 (2001); 
%
V. Sahni, M. Sami and T. Soura deep, Phys. Rev. {\bf D65}, 023518 (2002); 
%
J.S. Alcaniz, D. Jain, A. Dev, {\em ibid.}, {\bf 66}, 067301 (2002);
%
D. Jain, A. Dev, J.S. Alcaniz, {\em ibid.}, {\bf 66}, 083511 (2002);
%
Yu. Shtanov and V. Sahni, Class. Quant. Grav. {\bf 19}, L101 (2002);
%
V.~Sahni and Yu.~V.~Shtanov, Int. J. Mod. Phys. {\bf D11}, 1515 (2002); 
  JCAP, {\bf 0311}, 014 (2003);   
%
U.~Alam and V.~Sahni, ``{\em Supernova Constraints on Braneworld Dark Energy},"
  arXiv:astro-ph/0209443;
  %
K. Maeda, S. Mizuno, T. Torii, Phys. Rev.  {\bf D68}, 024033 (2003);
%
S. Nojiri and S. D. Odintsov, {\em ibid.}, {\bf 68}, 123512 (2003); 
%
R. G. Vishwakarma and P. Singh, Class. Quant. Grav. {\bf 20}, 2033 (2003);
%
Yu. Shtanov and V. Sahni, Phys. Lett.  {\bf B557}, 1 (2003);
%
{\em ibid.}, {\bf 599}, 137 (2004); 
%
R. Maartens, Living Rev. Rel. {\bf 7}, 7 (2004) [arXiv:gr-qc/0312059];
%
M. Sami and V. Sahni,  Phys. Rev. {\bf D70}, 083513 (2004); 
%
P. Brax, C. van de Bruck, A-C. Davis, Rept. Prog. Phys. {\bf 67}, 2183 (2004);
%
J.A.S. Lima, Braz. J. Phys. {\bf 34}, 194 (2004);
%
J.A. Alcaniz and N. Pires, Phys. Rev.  {\bf D70}, 047303 (2004);
%
A. Lue and G.D. Starkmann, {\em ibid.}, {\bf 70}, 101501 (2004);
%
A.~Padilla, ``{\em Infra-red modification of gravity from asymmetric branes},"
arXiv:hep-th/0410033; \ A.~Padilla, ``{\em Cosmic acceleration from asymmetric branes},"
arXiv:hep-th/0406157; V. Sahni and Yu. Shtanov, arXiv:astro-ph/0410221;
%
J.Ellis, N. E. Mavromatos, and M. Westmuckett, ``{\em Potentials between D-Branes 
in a Supersymmetric Model of Space-Time Foam}," arXiv:gr-qc/0501060;
%
K. Freese, ``{\em Cardassian Expansion: Dark Energy Density from Modified 
Friedmann Equations}," arXiv:astro-ph/0501675;
%
H.-B. Wen and X.-B. Huang, ``{\em Dark Energy Density in Brane World}," 
[arXiv:hep-th/0502078] Chin. Phys. Lett. {\bf 22}  816 (2005);
%
V. Sahni, ``{\em Cosmological Surprises from Braneworld models of Dark Energy},"
arXiv:astro-ph/0502032;
%
B. Cuadros-Melgar and E. Papantonopoulos, ``{\em The Need of Dark Energy for Dynamical 
Compactification of Extra Dimensions on the Brane}," arXiv:hep-th/0502169;
%
A. A. Andrianov, V.A.Andrianov, P. Giacconi, and R. Soldati,
``{\em Brane world generation by matter and gravity}," arXiv:hep-th/0503115;
%
A. S. Majumdar and N. Mukherjee, ``{\em Braneworld black holes in cosmology 
and astrophysics}," arXiv:astro-ph/0503473;
%
C.P. Burgess, D. Hoover, ``{\em UV Sensitivity in Supersymmetric Large Extra Dimensions: 
The Ricci-flat Case}," arXiv:hep-th/0504004;
%
A. Friaca, J.S. Alcaniz, and J. A. S. Lima, ``{\em An old quasar in a young dark 
energy-dominated universe?}" arXiv:astro-ph/0504031;
%
N. Bilic, G.B. Tupper, and R. D. Viollier, ``{\em Chaplygin-Kalb-Ramond Quartessence},"
arXiv:hep-th/0504082;
%
M. K. Mak, T. Harko, ``{\em Chaplygin gas dominated anisotropic brane world 
cosmological models}," arXiv:gr-qc/0505034.


\bibitem{DE}  A.~G.~Riess {\it et al.}  [Supernova Search Team Collaboration],
Astron.\ J.\  {\bf 116}, 1009 (1998);
S.~Perlmutter {\it et al.}  [Supernova Cosmology Project
Collaboration],
{\em ibid.},  {\bf 517}, 565 (1999);
A.~G.~Riess {\it et al.}  [Supernova Search Team Collaboration],
{\em ibid.},  {\bf 607}, 665 (2004);
C.~L.~Bennett {\it et al.}, 
Astrophys.\ J.\ Suppl.\  {\bf 148}, 1 (2003); 
D.~N.~Spergel {\it et al.}  [WMAP Collaboration], 
{\em ibid.},  {\bf 148}, 175 (2003); 
M.~Tegmark {\it et al.}  [SDSS Collaboration],
Phys.\ Rev.\ D {\bf 69}, 103501 (2004);
K.~Abazajian {\it et al.}, arXiv:astro-ph/0410239;
K.~Abazajian {\it et al.}  [SDSS Collaboration],
Astron.\ J.\  {\bf 128}, 502 (2004);
K.~Abazajian {\it et al.}  [SDSS Collaboration],
{\em ibid.},  {\bf 126}, 2081 (2003);
E.~Hawkins {\it et al.},
Mon.\ Not.\ Roy.\ Astron.\ Soc.\  {\bf 346}, 78 (2003);
L.~Verde {\it et al.},
{\em ibid.},  {\bf 335}, 432 (2002).

\bibitem{Note} We only consider here the case in which the extra
 dimensions are all spacelike. It is not difficult to show that our main
 conclusions will also apply to those cases in which the signature of the
 extra dimensions is arbitrary.

\bibitem{Taub80} A.H. Taub, J. Math. Phys. {\bf 21}, 1423 (1980).


\bibitem {Chandra83}  S. Chandrasekhar, {\em The Mathematical Theory 
of Black Holes} (Clarendon Press, Oxford University Press, Oxford, 1983),
pp.66-68.


\bibitem{Bernstein} D. L. Bernstein, {\it Existence Theorems in
 Partial Differential Equations} (Princeton, Princeton University
 Press, 1965), pp.109-111.

\bibitem{KST88} C. A. Kolassis, N.O. Santos, and D. Tsoubelis, Class.
 Quantum Grav.  {\bf 5}, 1329 (1988).

\bibitem{HE73} S.W. Hawking and G.F.R. Ellis, {\em The Large Scale
Structure of Spacetime} (Cambridge University Press, Cambridge, 1973).

\bibitem{Zhou92} Z.-H. Zhou, Helv. Phys. Acta {\bf 65}, 767 (1992).

\bibitem{Israel77} W. Israel, Phys. Rev. {\bf D15}, 935 (1977).

\bibitem{PW00} A.Y. Miguelote, M.F.A. da Silva, A.Z. Wang, and N.O. Santos,
 Class. Quantum Grav. {\bf 18}, 4569 (2001), and references therein.

\bibitem{Gr00} R. Gregory, Phys. Rev. Lett. {\bf 84}, 2564 (2000); T.
 Gherghetta and M. Shaposhnikov, {\em ibid.}, {\bf 85}, 240 (2000); M.
 Giovannini, H. Meyer, and M. Shaposhnikov, Nucl. Phys. {\bf B619}, 615
 (2001); P. Bostock, R. Gregory, I. Navarro, and J. Santiago, 
Phys. Rev. Lett. {\bf 92}, 221601 (2004);  and references therein.

\bibitem{Zhuk} V. Sahni, ``{\em Cosmological surprises from Braneworld
models of Dark Energy}," arXiv:astro-ph/0502032; M. Bouhmadi-L\'opez, 
P.F. Gonzalez-Diaz, and A. Zhuk, Class. Quantum Grav. {\bf 19}, 4863 (2002); 
``{\em Perfect Fluid brane-world model,}" arXiv:hep-th/0111291 (2001);
and references therein.

\end{thebibliography}
\end{document}